# Liquid crystal phases with unusual structures and physical properties formed by acute-angle bent-core molecules

Bing-Xiang Li,[1,2,†] Yuriy A. Nastishin,[1,3,†] Hao Wang,[1,2] Min Gao,[1] Sathyanarayana Paladugu,[1] Ruipeng Li,[4] Masafumi Fukuto,[4] Quan Li,[1,2] Sergiy V. Shiyanovskii,[1,2] and Oleg D. Lavrentovich[1,2,5,*]

[1]Advanced Materials and Liquid Crystal Institute, Kent State University, Kent, OH, 44242, USA
[2] Materials Science Graduate Program, Kent State University, Kent, OH, 44242, USA
[3]Hetman Petro Sahaidachnyi National Army Academy, 32, Heroes of Maidan street, Lviv, 79012, Ukraine
[4]National Light Source II, Brookhaven National Laboratory, Upton, New York 11973, USA
[5]Department of Physics, Kent State University, Kent, OH, 44242, USA
*olavrent@kent.edu
[†] B. L. and Y. N. contribute equally to this work.

## Abstract

Liquid crystals formed by acute-angle bent-core (ABC) molecules with a 1,7 naphthalene central core show an intriguing phase behavior with the nematic phase accompanied by poorly understood additional phases. In this work, we characterize the physical properties of an ABC material, such as birefringence, dielectric permittivities, elastic constants, and surface alignment and present an X-ray diffraction and transmission electron microscopy studies of their ordering. The ABC molecular shape resembling the letter $\lambda$ yields a very small splay elastic constant in the uniaxial nematic phase and results in the formation of a tetragonal positionally ordered columnar phase consisting of molecular columns with a uniform uniaxial director that can be bent but not splayed.

# I. INTRODUCTION

The molecular shape is a key factor responsible for physical properties of liquid crystals. The simplest rod-like molecules produce a uniaxial nematic (N), widely used in applications. Lately, there has been much interest in the so-called bent-core molecules, formed by two rod-like segments attached to each other at some angle $\beta$. This kinked shape leads to fundamentally new properties and phases, as reviewed recently [1, 2]. Nematics formed by molecules with $110^\circ \leq \beta \leq 170^\circ$ often exhibit a bend elastic constant that is smaller than the splay constant, $K_{33} < K_{11}$, [1-8], which is opposite to the trend $K_{33} > K_{11}$ found in conventional rod-like nematics. The tendency to bend might be so strong that the bent-core materials exhibit new states, such as the twist-bend nematic [9-13] or a heliconical cholesteric [14]. Most of the studies so far focused on the obtuse angle bent-core (OBC) molecules, $\beta > 90°$. Mesomorphism of the acute-angle bent-core (ABC) molecules with $\beta < 90°$ is poorly understood. Kang, Watanabe and their colleagues [15-18] brought about interesting results for ABC molecules such as the one in Fig.1. The molecule is formed by two arms attached in an asymmetric fashion to the central 1,7-naphthalene core. It produces a high temperature uniaxial N and a low-temperature phase, identified as a tetragonal packing of long cylinders. It is claimed [15-18] that each cylinder contains a double-twisted director field which is necessarily associated with a bend. The cylinders are of alternating chirality: a left-twisted cylinder has four close neighbors of right twisted cylinders and vice versa.

Recent studies demonstrate that surprising new types of the nematic ordering can be formed even by seemingly simple rod-like molecules, such as 4-[(4-nitrophenoxy)carbonyl]phenyl2,4-dimethoxybenzoate, abbreviated RM734. Mandle, Cowling and Goodby found that these materials could exhibit a transition from a uniaxial nematic to some other nematic upon cooling [19, 20]. In studies that followed, Mertelj et al [21, 22] suggested that the transition is to a splay nematic, in which the director $\hat{\mathbf{n}}$ experiences a periodic splay with a period $5-10$ μm [23]. The spontaneous splay was attributed to the wedge shape of the molecules and by an ensuing smallness of $K_{11}$ [21]. The situation is similar to the formation of a twist-bend nematic, which is heralded by a decrease of $K_{33}$ [13, 24, 25]. The overall structure of the splay nematic was suggested to be of an antiferroelectric type [21]. Clark et al [26] re-examined the electro-optical and textural properties of RM734 and concluded that the material is in fact a three dimensional ferroelectric uniaxial nematic in which the large longitudinal dipole moments (about 11 D) are parallel to each other. A unidirectional ferroelectric order has been also announced by Kikuchi et al in 2017 [27] for a different compound with dipole moments $\sim 9.4$ D.

Motivated by these recent advancements and by the need to establish a structure-property relationship for ABC compounds, in this work, we characterize mesomorphic properties of the liquid crystal 1,7-naphthylene bis(4-(3-chloro-4-(4-(hexyloxy)benzoyloxy)phenyliminomethyl) benzoate), abbreviated as 1Cl-N(1,7)-O6, Fig. 1; see Supplemental Material for the chemical synthesis details. The material shows a uniaxial N phase at high temperatures and transforms into a columnar $Col_t$ phase of tetragonal symmetry at lower temperatures. $K_{11}$ in the N phase is smaller than the twist elastic constant $K_{22}$ by a factor of 2 and smaller than $K_{33}$ by a factor of 6. We explain this anomalous elasticity by the alignment of the longitudinal axes of $\lambda$-shaped molecules parallel to the director. The splay is facilitated by flip-flops of the molecules. Using the X-ray diffraction data, we propose that the $Col_t$ phase is formed by columnar aggregates of a diameter roughly equals to the distance between the legs of a single $\lambda$- molecule. The columns are of two types, one polar and one apolar. In the polar columns, all $\lambda$- molecules point along the same direction in a ferroelectric fashion, either upward or downward along the column. In the apolar aggregates, one half of the molecules points upward and the other half points downward. Each polar column is neighbored by four columns of opposite polarity, which yields a tetragonal lattice. The apolar columns fill gaps between the polar columns.

## II. MATERIALS AND METHODS

The 1Cl-N(1,7)-O6 molecule in Fig. 1 is of a $\lambda$-shape, with two legs being of slightly different length. Numerical simulations, Fig.1b, suggest that the legs tilt away from the naphthalene core plane, to minimize the conformation energy.

The optical and surface anchoring properties of the material 1Cl-N(1,7)-O6 are explored by polarizing optical microscopy (POM), PolScope observations, and by measurements of the birefringence. Dielectric properties are examined in the frequency range from 100 Hz to 1 MHz using LCR-meter (4284A, Hewlett-Packard) operated by a Labview program. The temperature of the cells is controlled by a LTS350 hot stage and a TMS-94 controller (both from Linkam Scientific Instruments). The dielectric permittivity $\varepsilon_\parallel$ parallel to the director is measured in a cell of thickness $15.5\,\mu m$ with substrates covered with a surfactant hexadecyltrimethylammonium bromide (HTAB). The overlapped area of two square-patterned indium tin oxide (ITO) electrodes is $1\,cm \times 1\,cm$. The perpendicular component $\varepsilon_\perp$ is measured using planar cells with SE-1211 coatings. The cells' gap is fixed by glass microspheres of calibrated diameter and measured by recording the light interference spectrum of an empty cell. The cells are filled by capillary action at $T = 190°C$. $K_{11}$ and $K_{33}$ are measured in planar cell of thickness $d = 5.8\,\mu m$ with the aligning polyimide SE-1211 layers. $K_{11}$ is determined from the threshold voltage of the splay Frederiks transition and $K_{33}$ is determined by fitting $C(U)$ well above the threshold, where splay is accompanied by bend [28]. To measure $K_{22}$, we add 2.8 wt% of a chiral dopant S-811 to 1Cl-N(1,7)-O6, and then use the electric field to unwind the resulting cholesteric helix in a homeotropic cell with SE-5661 alignment layers; the voltage at which the helix is unwould defines $K_{22}$ [29].

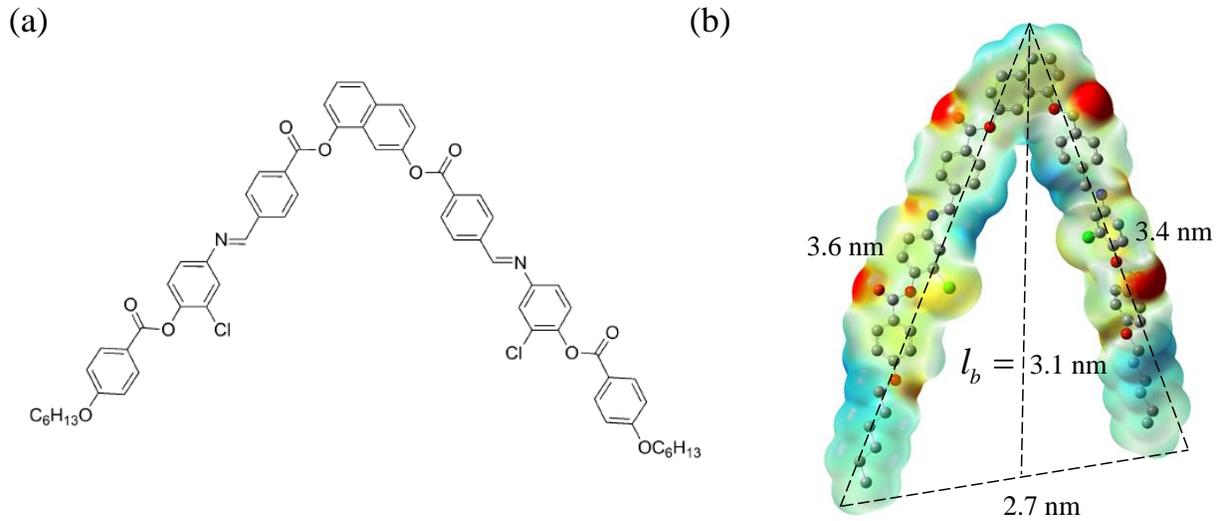

FIG. 1. (a) Molecular formula and (b) one of the conformations of the studied ABC material 1Cl-N(1,7)-O6. Part (b) obtained by numerical simulation using software Gaussian 09W to find a minimum energy of a conformation based on density functional theory (DFT) at B3LYP 6-31G(d) level.

The Freeze-fracture (FF) Transmission Electron Microscopy (TEM) and cryo-TEM of the samples follow the procedures described previously [30]. In FFTEM, the samples are heated into the N phase and equilibrated at $190°C$. After that, they are slowly cooled down at a rate of $1°C/min$ to the desired temperatures and equilibrated for 5 min. The sample is then plunge frozen in liquid nitrogen and fractured at $-165°\text{C}$. Replicas are prepared by depositing thin ($4\,\text{nm}$) layers of Pt/C at a $45°$ angle, followed by $20\,\text{nm}$ C film deposition at normal incidence. Unlike FFTEM, cryo-TEM produces direct images of ultrathin specimens. We use thin (up to $20\,\text{nm}$) films of the material supported by a holey carbon film. For cryo-TEM observation, the material supported by the carbon film is heated to the N phase and slowly cooled down to the desired temperature. The equilibrated samples are then rapidly frozen in liquid ethane and observed under the TEM.

X-ray experiments are performed the 11-BM CMS beamline at National Synchrotron Light Source II, Brookhaven National Laboratory. The beamline is configured for a collimated X-ray beam with a beam size of cross-section $0.2\,\text{mm} \times 0.2\,\text{mm}$ and divergence of $0.1\,\text{mrad} \times 0.1\,\text{mrad}$ with the X-ray energy of $17\,\text{keV}$. Two detectors, Pilatus 2M and Photon Science CCD, are placed at $3\,\text{m}$ and $236\,\text{mm}$ away from samples, respectively, to collect the SAXS and WAXS signals. The sample-to-detector distances were calibrated by using a silver behenate standard. The experiments were performed with the in-situ thermal control by using an Instec HCS402 hot stage with the temperature deviation less than $0.1°\text{C}$. Two types of samples are studied. The first represents a quartz capillary of an inner diameter 1 mm mounted into an aluminum cassette between two $1\,\text{T}$ magnets. The second sample represents a flat cell formed by two glass plates with transparent ITO electrodes. Each glass substrate is $170\,\upmu\text{m}$ thick and the cell gap is also $170\,\upmu\text{m}$. Using a waveform generator (Stanford Research Systems, Model DS345) and a power amplifier (Krohn-hite Corporation, Model 7602), we apply a sinusoidal voltage of the amplitude $100\,\text{V}$ at the frequency $2\,\text{kHz}$, so that the director aligns along the field and perpendicular to the glass plates. The cell is mounted for transmission X-ray scattering with the incident beam normal to the plates, such that the director points along the beam.

## III. EXPERIMENTAL RESULTS
### A. Phase diagram

The phase diagram obtained by POM textural observations is shown in Fig. 2, which is consistent with the DSC data in Ref. [18]. The POM textures of different phases are shown in Figs.

3-6. On heating, the material forms a uniaxial N phase between 183°C and $T_{NI} = 215$°C, where $T_{NI}$ is the N-to-isotropic (I) transition point. On cooling, the I-N transition occurs through nucleation of spherical droplets, Fig. 3. Each droplet shows a texture of a Maltese cross with four dark brushes and a point defect in the center, Fig. 3a,b,c. Conventional POM with two crossed linear polarizers does not allow one to establish unequivocally whether the Maltese cross is caused by a radial configuration of the director or by a concentric configuration, since both these configurations would yield four dark brushes in the regions where the director is parallel to the polarizer (P) and analyzer (A). To establish the director orientation around the defect core we use a full wavelength plate (FWP), also called a red-plate optical compensator. The FWP is a slab of a uniaxial crystal of positive birefringence that produces a 530 nm optical phase retardance. In POM observations, the FWP is placed between the crossed P and A with its optic axis along the angle bisector of quadrants I and III formed by P and A. In absence of any birefringent sample, it produces the red-violet interference color, which separates the first- and second-order interference colors on the Michel-Lévy color interference chart. When a birefringent sample is present, the red-violet color remains in the region in which the sample's optic axis (the director in our case) is along the P and A directions. Other orientations of the director produce either a blue-green or yellow-orange color, depending on whether the additional phase retardance adds to the phase retardance of FWP or subtracts from it. This property is used either for the determination of the sign of optical birefringence of an optically anisotropic plate if the direction of the optic axis is known or for the determination of the direction of the optic axis if the optical sign is known [31]. As will be demonstrated in section D below, the optical birefringence of the liquid crystalline phases of the explored material is positive. Observation of the Maltese textures of the N droplets with the FWP shows that the I and III quadrants acquire a blue interference color, while the II and IV quadrants become yellow, Fig.3 e-g. Therefore, the director in the N droplets is perpendicular to the N-I interface and forms a radial point defect-hedgehog [32], Fig. 3e, with predominantly splay deformation of the director. Radial director field in the N nuclei and homeotropic alignment at the N-I interface are unusual for thermotropic nematics, since the director orientation at the N-I interface in many of them is tilted, as is the case of cyanobiphenyls [33]. The nuclei grow and coalesce, Fig. 3b,c. When they touch the top and bottom glass plates, the director realigns perpendicularly to the plates, Fig. 3b-d.

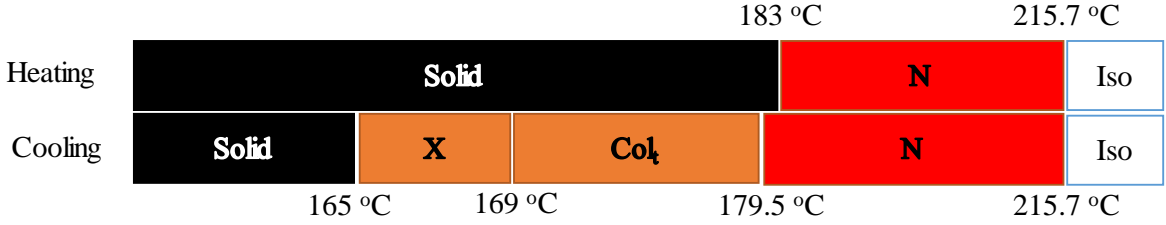

FIG. 2. Phase diagram of 1Cl-N(1,7)-O6. Heating rate is 1°C/min for heating and cooling rate is 0.5°C/min for cooling.

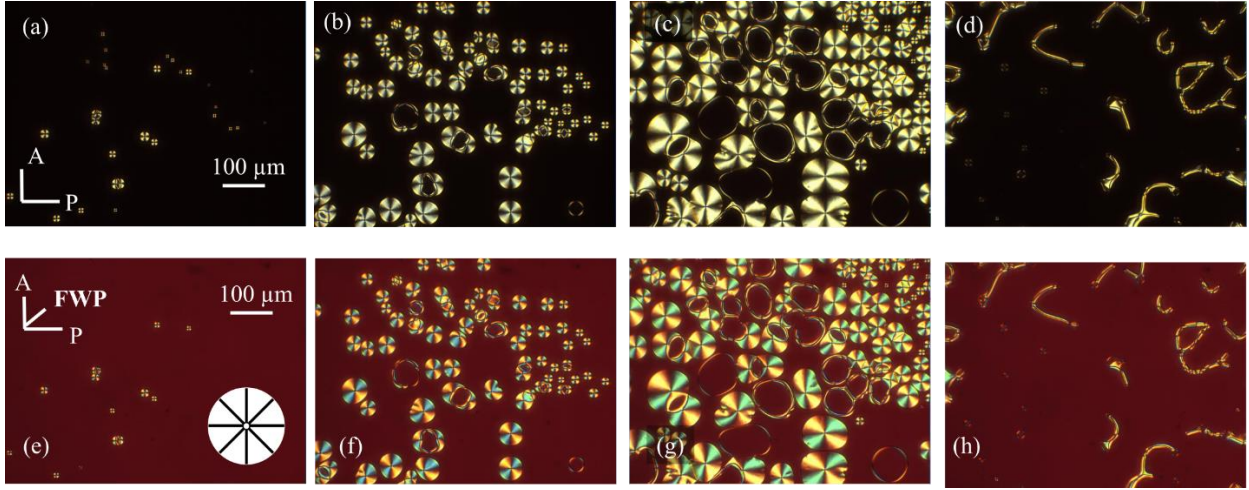

FIG. 3. POM of (a) nucleation and growth (b,c and e,f) of N droplets from the isotropic melt; director shows splay deformations; (c,d and g,h) nuclei that touch both glass plates adopt homeotropic alignment of the director. The slab is confined between two clean glasses with no alignment layers. Textures (a-d) are viewed between two crossed polarizers; textures (e-h) are viewed with a full wavelength optical compensator of retardance 530 nm with the slow axis along the line "FWP". The temperature is reduced from 214.7°C in (a) to 213.5°C in part (d).

On cooling, the N phase exists down to $T_{NCol_t} = 179.5°C$. Between $T_{NCol_t}$ and the solidification temperature $T_S = 165°C$, one observes two other phases, a higher-temperature $Col_t$ and a lower-temperature X phase. Figure 4 demonstrates a textural transition from N to $Col_t$ and then to X. The temperature $T_{Col_t X}$ of the $Col_t$-to-X transition depends on the sample prehistory. If a fresh sample is cooled for the first time with a rate (0.5°C/min), $T_{Col_t X} = 169°C$. However, if the sample is kept for several hours at elevated temperatures, either in the N or in the $Col_t$ phase, the material degrades irreversibly, and the temperatures of all transitions shift noticeably. Namely, $T_{NI}$ and $T_{NCol_t}$ decrease while $T_{Col_tX}$ increases, approaching $T_{NCol_t}$ by approximately $1°C$ per hour of sample stabilization at any temperature above $169°C$.

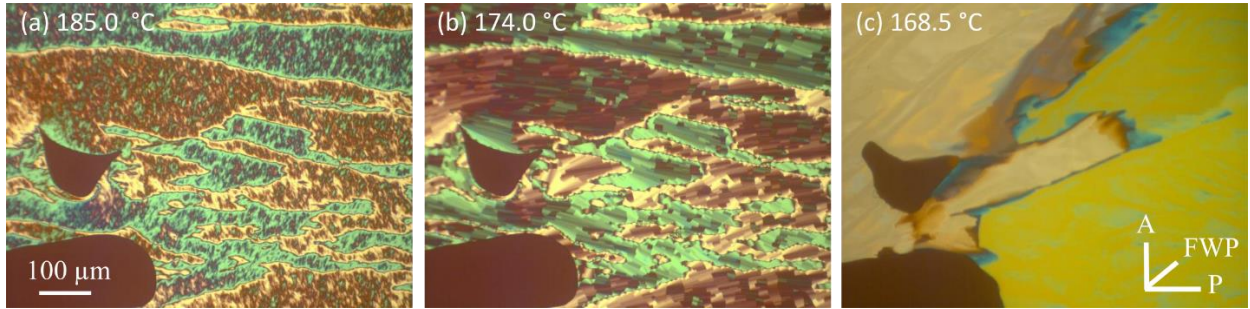

FIG. 4. Textural transition upon cooling from (a) the N phase; the cell with homeotropic alignment (polyimide PI-2555 coating) is sheared; (b) $Col_t$ phase with a texture reminiscent of columnar phases texture; (c) X phase. POM observation with FWP.

## B. Alignment

Table 1 lists the alignment effect of seven different materials on the N phase of 1Cl-N(1,7)-O6 and a conventional rod-like nematic pentylcyanobiphenyl (5CB). The used alignment materials are four polyimides, PI 2555 (HD MicroSystems), SE 1211 (Nissan Chemicals), SE 5661 (Nissan Chemicals), SE 7511 (Nissan Chemicals), two surfactants from Sigma-Aldrich, hexadecyl trimethyl ammonium bromide (HTAB), dimethyloctadecyl [3-(trimethoxysilyl)propyl] ammonium chloride (DMOAP), and clean glass. All but one produce homeotropic orientation of 1Cl-N(1,7)-O6, with the director perpendicular to the bounding plates, $\hat{\mathbf{n}} = (0,0,1)$. Rubbing the homeotropic substrates leads to the tilt of the nematic director by an angle $\sim 10-15°$. Homeotropic alignment is produced even by the polyimide PI2555, which is widely used for planar orientation of the conventional nematics such as 5CB. Homeotropic alignment by clean glass can be related to the hydrophilic nature of the substrate; many thermotropic nematics align similarly on clean glass [34]. The polyimid SE-1211 shows multiple regimes of alignment. If the nematic phase of 1Cl-N(1,7)-O6 is obtained by cooling down from the I phase, SE-1211 imparts a homeotropic alignment, within about $1°C$ from the I--N transition point. At lower temperatures, the alignment becomes tilted, as evidenced by the defects of strength $+1$ and $-1$, each with four extinction brushes, Fig. 5. Isolated semi-integer defects are prohibited at the tilted director alignment since the director projection onto the plates is a vector rather than a director; therefore, only integer strength defects could be observed [31]. If the material is filled into the cell in the N phase, the resulting alignment by SE-1211 is tangential, showing Schlieren textures with isolated $+1/2$ and $-1/2$ disclinations. The reason is as follows. For strictly tangential boundary conditions, the surface orientations $\hat{n}$ and $-\hat{n}$ are not distinguishable from each other. Therefore, not only integer but also semi-integer defects are topologically permissible [31]. If the director is even slightly tilted away from the surface, the tilts "up" and "down" are no longer indistinguishable: the "up" and "down" directors placed in the same point form a letter "x". Therefore, if any semi-integer defect existed before the tilt, it would give rise to a wall defect that starts at the defect core and provides a reorientation from the "up" to "down" state through a strictly tangential director. Such a wall is energetically unfavorable. This is why under strictly tangential anchoring one could observe semi-integer defects-disclinations, while under titled conical boundary conditions the defects are of an integer strength, Fig.5; these are typically surface point defects-boojums [32]. For more details and for illustration of this effect, see Ref. [31], pages 395-396.

Table 1. Effect of different surface agents on the alignment of ABC nematic 1Cl-N(1,7)-O6 and rod-like nematic pentylcyanobiphenyl (5CB).

| Alignment agent | ABC nematic | 5CB |
|---|---|---|
| PI 2555 | Homeotropic | Tangential |
| SE 1211 | Homeotropic, Tangential or Tilted | Homeotropic |
| SE 5661 | Homeotropic | Homeotropic |
| SE 7511 | Homeotropic | Homeotropic |
| HTAB | Homeotropic | Homeotropic |
| DMOAP | Homeotropic | Homeotropic |
| Clean glass | Homeotropic | Homeotropic |

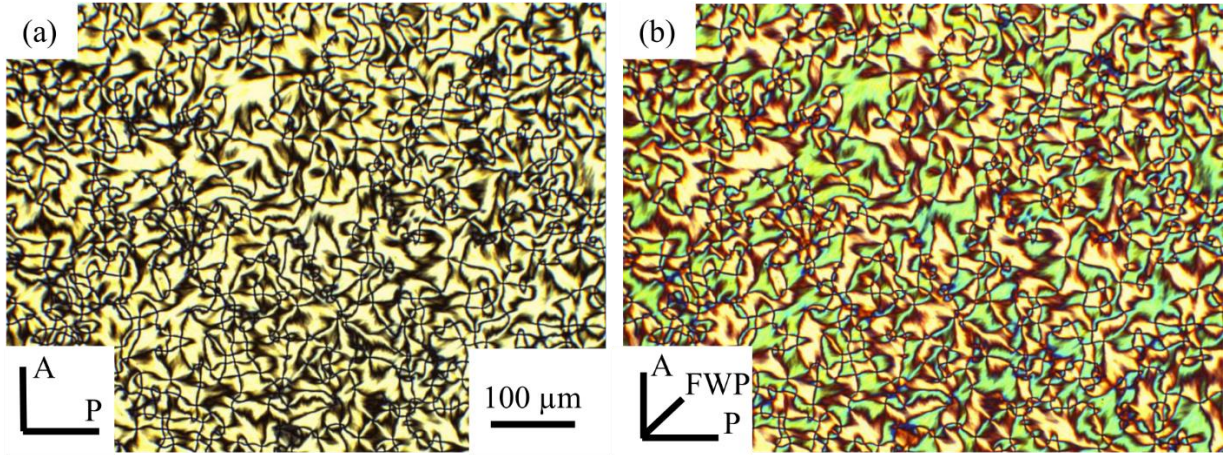

FIG. 5. Nematic Schlieren texture ($T = 210°C$) in a cell with SE1211 alignment layers, filled at the temperature $T = 220°C$ in the I phase. Optical microscopy with two crossed polarizers without (a) and with (b) a full wave plate. Note presence of defect centers with four dark brushes and absence of defect centers with two dark brushes. The feature is caused by presence of integer-strength defects and absence of semi-integer defects, demonstrating a tilted conical alignment of the director at the bounding plates with SE1211 at this temperature.

In free-standing films, prepared in the openings of the cooper electron microscopy grids with $(40 \times 40)\,\mu m^2$ square openings, the N-air interface imposes tangential orientation, as evidenced by observations of pairs of $1/2$ defects, Fig. 6a, [31, 35-37]. The director field around the cores of the $1/2$ disclinations reveals two features: (1) there are no defect walls attached to the cores of disclinations, which indicates that the director is strictly parallel to the N-air interface; (2) the main in-plane director deformation is splay, signaling $K_{11} \ll K_{33}$. In contrast, the textures of Col$_t$ phase show bend rather than splay, which is a common feature of the columnar phases, Fig. 6b.

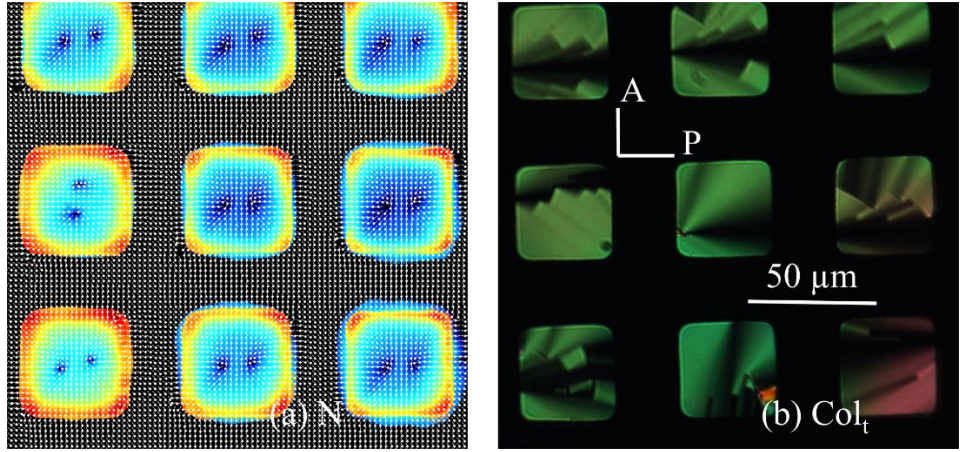

FIG. 6. Freely suspended films of 1Cl-N(1,7)-O6 (a) PolScope texture of N phase with $1/2$ defects and predominant splay of the director; $T = 190°C$; (b) POM texture of $Col_t$ that reveals bend of the optic axis; one region of bend is traced with an arch in the central square. $T = 176°C$.

## C. Dielectric permittivity

Frequency dispersions for the parallel $\varepsilon_\parallel$ and perpendicular $\varepsilon_\perp$ components of the dielectric permittivity are shown in FIG. 7a. The material is of a dual-frequency type, as $\Delta\varepsilon = \varepsilon_\parallel - \varepsilon_\perp > 0$ at low frequencies, $f < f_c$ and $\Delta\varepsilon < 0$ at high frequencies, $f > f_c$, Fig. 7b. The crossover frequency $f_c$ is in the range $(10-60)\,\text{kHz}$, depending on the temperature. The temperature dependencies $\varepsilon_\parallel(T)$ and $\varepsilon_\perp(T)$ measured at $f = 1\,\text{kHz}$ are relatively weak, Fig. 7c. Both permittivities show a slight increase below the N-$Col_t$ transition, Fig. 7c. The value of $\Delta\varepsilon$ in the N phase is relatively small, which correlates with the apolar character of the N phase.

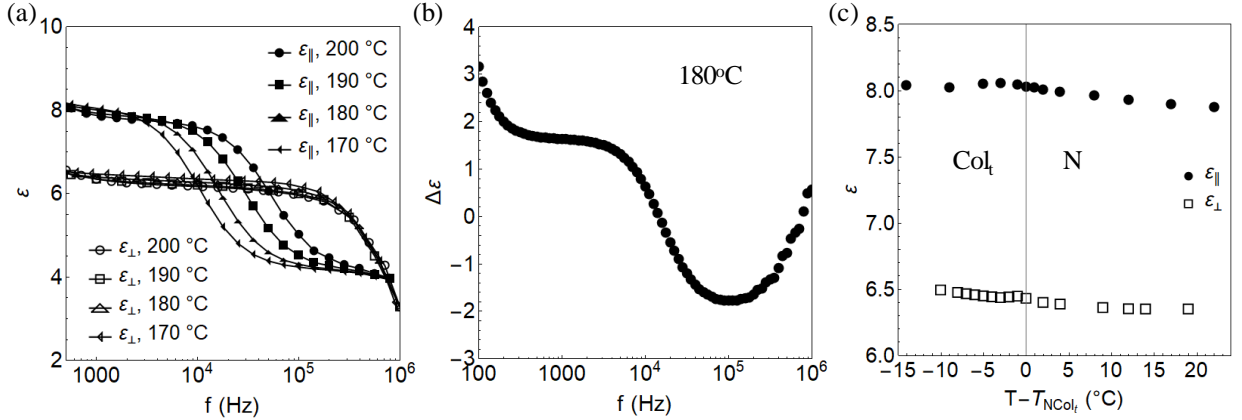

FIG. 7. (a) Frequency dispersion of parallel and perpendicular components of dielectric permittivity of 1Cl-N(1,7)-O6 at different temperatures. (b) Frequency dispersion of the dielectric anisotropy in the nematic phase at $T = 180°C$; (c) Temperature dependencies $\varepsilon_\parallel(T)$ and $\varepsilon_\perp(T)$ measured at $f = 1\,\text{kHz}$ for 1Cl-N(1,7)-O6 in N and Col$_t$ phases.

### D. Birefringence

Optical conoscopy of homeotropic N cells with the director along the normal to the bounding plates, $\hat{\mathbf{n}} = (0,0,1)$, shows a Maltese cross with dark brushes oriented along the directions of crossed polarizers, Fig.8a, which means that the N phase is optically uniaxial. Conoscopic observation of the sample is performed by using a cone of converging rays of light. The homeotropic alignment of $\hat{\mathbf{n}} = (0,0,1)$ observed by a converging cone of rays appears as a radial configuration of the effective optical axis converging in the center of the conoscopic pattern in Fig. 8a. To establish the sign of birefringence, an FWP with 530 nm retardance is inserted with the optic axis along the angle bisector of the I and III quadrants formed by the polarizers P and A, Fig.8b. The POM texture in Fig.8b shows that the I and the III quadrants acquire a blue interference color when the FWP is inserted. It implies that in these quadrants, the phase retardance of the sample adds to the phase retardance of the FWP, i.e., that the birefringence of the liquid crystal is positive. Accordingly, the quadrants II and IV become yellow, since the optical axes of the FWP and the liquid crystals are orthogonal to each other and the overall phase retardance is reduced. The texture in Fig.8b thus establishes that the birefringence of the N phase is positive, $\Delta n = n_e - n_o > 0$, where $n_e$ and $n_o$ are the extraordinary and ordinary refractive indices, respectively.

For an optically negative N, the I and III quadrants would be yellow, and the II and IV quadrants would be blue.

When the sample is cooled to the $Col_t$ phase, the Maltese cross shifts towards the periphery of the field of view, Fig. 8c, indicating that the optic axis tilts by about, from the normal to the bounding plates. The tilt angle is $\theta \approx 20°$, as calculated from the relationship $NA \cdot \sin\theta = R/R_0$, where $NA = 0.65$ is the numerical aperture of the objective, $R_0$ is the radius of the circular conoscopy pattern, and $R$ is the shift of the cross. In Fig.8c, $R/R_0 \approx 0.22$. $Col_t$ is also of a positive birefringence, $\Delta n > 0$, as follows from the observations with the FWP, Fig. 8d.

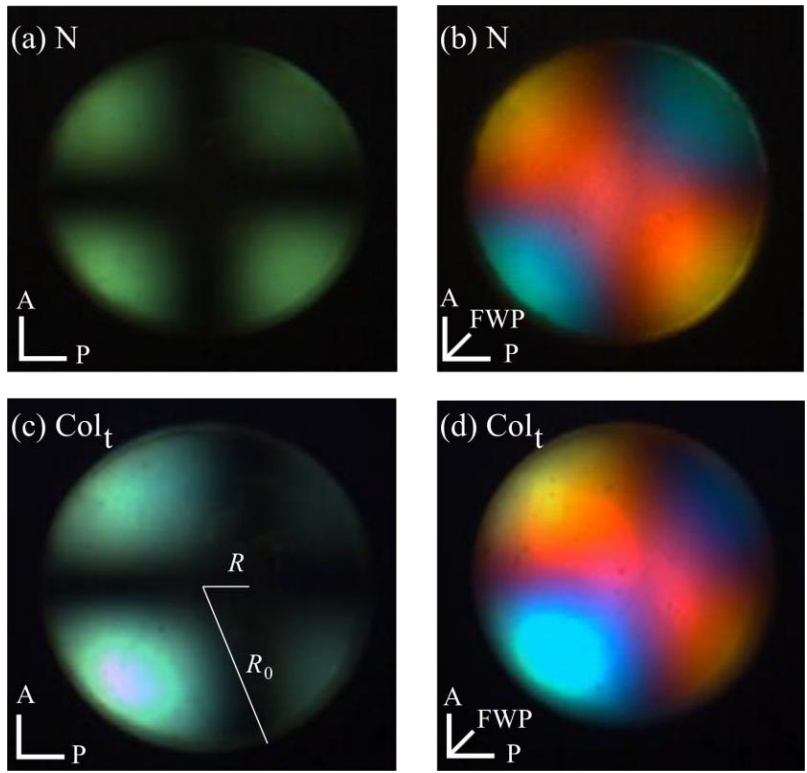

FIG. 8. Conoscopic patterns of the homeotropic cells with N (a,b) and $Col_t$ (c,d) phases of 1Cl-N(1,7)-O6; observations with two crossed polarizers in (a,c) and with an additional FWP in (b,d).

The temperature dependence of birefringence at the wavelength 546 nm was measured using PolScope and a planar cell of thickness 1.9 μm aligned by a unidirectionally rubbed SE-1211, Fig. 9. The measured values are close to those known for conventional nematics formed by rod-like molecules.

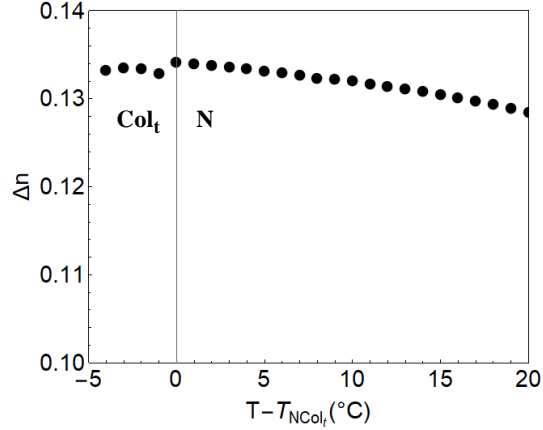

FIG. 9. Temperature dependence of birefringence at the wavelength 546 nm of 1Cl-N(1,7)-O6 in N and $Col_t$ phases.

### E. Frederiks transitions and elastic constants in the N phase

The temperature dependencies of the elastic constants are shown in Fig. 10 a. The most unusual result is that the splay elastic constant $K_{11}$ is much smaller than the twist modulus $K_{22}$ (by a factor $\sim 1/2$) and the bend modulus $K_{33}$ (by a factor of $\sim 1/6$), Fig. 10 a. The result is also supported by a qualitative feature in Fig.10 b: the Schlieren N textures show a clear predominance of splay deformations over bend in the structure of +1 defects, which illustrates clearly that $K_{11}$ is much smaller than $K_{33}$. To the best of our knowledge, this is the first report on an anomalously small splay elastic constant in ABC material. Previously, Lee et al [38] reported that ABC molecules, added in a weight proportion up to 10 wt%, decrease the splay constant of a rod-like nematic from 13 pN to 3 pN. A very small $K_{11} \approx 1\,\text{pN}$ is also reported for the uniaxial N phase of RM734 by Mertelj et al [21].

The Frederiks effect in 1Cl-N(1,7)-O6 can be triggered by a magnetic field. For a homeotropic cells of thickness $30\,\mu\text{m}$, the threshold field is $0.26\,\text{T}$ at 203°C, which yields $K_{33}/\Delta\chi = 4.8\times 10^{-6}\,\text{N}$, where $\Delta\chi = \chi_\parallel - \chi_\perp > 0$ is the diamagnetic anisotropy of 1Cl-N(1,7)-O6. At the same temperature, the electric Frederiks transition yields $K_{33} = 8\,\text{pN}$; thus at 203°C, $\Delta\chi = 1.7\times 10^{-6}$, which is comparable to the diamagnetic anisotropy of rod-like mesogens such as 5CB.

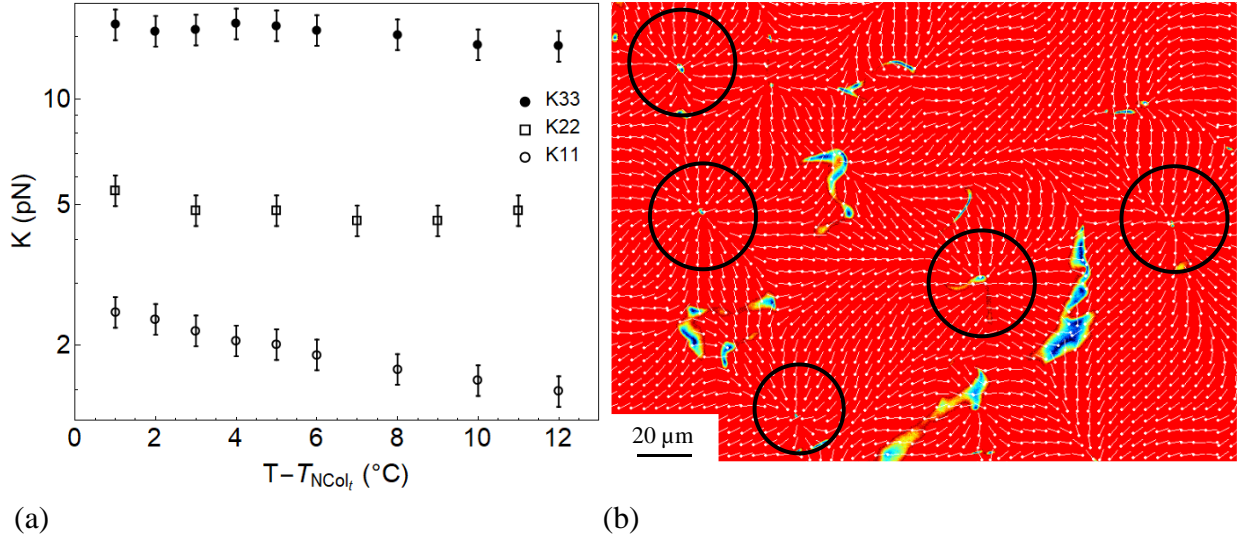

(a)                                      (b)

FIG. 10. (a) Temperature dependencies of the elastic constants in the N phase of 1Cl-N(1,7)-O6; (b) PolScope Schlieren N texture of 1Cl-N(1,7)-O6 demonstrating prevalence of splay over bend around +1 defects of radial aster geometry marked by open circles.

### F. Frederiks transitions in Col$_t$ phase

**Planar to homeotropic transition**. A high voltage $U > U_{th}$, where $U_{th} \approx 24\,\text{V}$, of the frequency $f = 1\,\text{kHz}$, at which $\Delta\varepsilon > 0$, reorients the director from planar to homeotropic state in a planar $6.0\,\mu\text{m}$ cell. Voltage dependence of optical retardance for the Col$_t$ phase, Supplementary Fig. S1a, shows behavior reminiscent of the Frederiks transition in nematics, Supplementary Fig. S1b. However, in the N phase, the threshold is much lower, $U_{th} \approx 1\,\text{V}$. If one assumes that the threshold in Col$_t$ depends on the splay modulus, $U_{th} = \pi\sqrt{K_{11}/\varepsilon_0\Delta\varepsilon}$, then with the known $\Delta\varepsilon$ one obtains a very rough estimate $K_{11} = \varepsilon_0\Delta\varepsilon U_{th}^2/\pi^2 \approx 10^3\,\text{N}$, 500 times larger than in the N phase.

**Homeotropic to planar transition**. To stabilize the homeotropic Col$_t$ state, the cell is cooled in the presence of an electric field with $f = 1\,\text{kHz}$, at which $\Delta\varepsilon > 0$. The homeotropic state could be switched into a planar state by an electric field at $f = 200\,\text{kHz}$, at which $\Delta\varepsilon < 0$. The reorientation involves nucleation of domains resembling developable domains of columnar phases [31], Fig. 11. The transition is of the first order (as evidenced by nucleation and by the sharp interface between the two states), i.e., dramatically different from the second-order Frederiks transitions in the N phase. The bent interface between the homeotropic and the deformed region together with the POM and Polscope textures that show the director parallel to the interface, suggest that the deformation of the director is of a bend type.

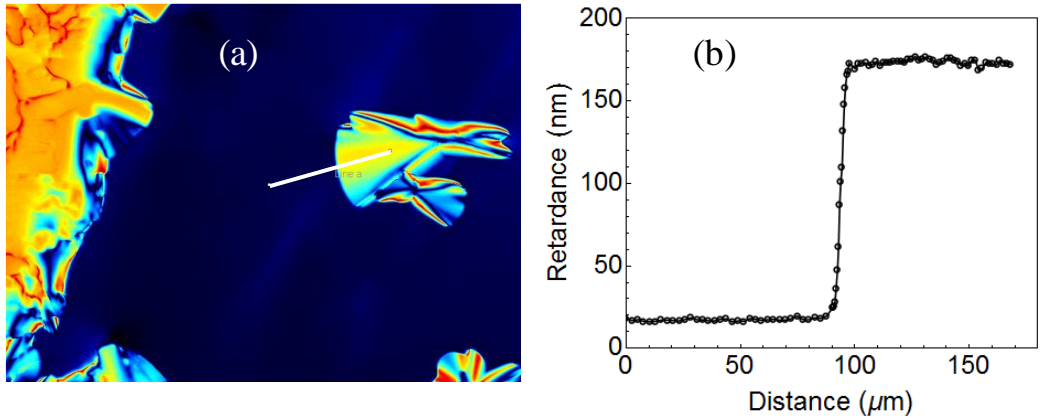

FIG. 11. (a) PolScope texture of the Frederiks transition in a homeotropic $Col_t$ cell of 1Cl-N(1,7)-O6; $U = 24\text{ V}$; $f = 200\text{ kHz}$; (b) optical retardance across the interface between the realigned material and the homeotropic region along the white line in part (a). Temperature $T = T_{NCol_t} - 4°C$.

The Frederiks transition in N can be of the first order only under very special circumstances, such as a strong difference in the values of splay and bend Frank constants and peculiar types of surface anchoring [39-41]. On the other hand, layered and columnar phases in which either bend or splay deformations are prohibited by the requirement of equidistant positional order of layers or columns, routinely show the first-order Frederiks transitions with nucleation of domains such as focal conics [13, 42, 43].

### G. Freeze-fracture and cryo-TEM of the material in the $Col_t$ phase

The freeze-fracture TEM textures, Fig.12, and cryo-TEM textures, Fig.13, prove periodic ordering of $Col_t$ but do not establish the predominant period since it varies broadly from sample to sample and even within the same sample. The FFTEM texture of the $Col_t$ phase quenched from 179°C in Fig. 12a exhibits linear periodic structures with a period $P$ in the range $(7-8)\text{ nm}$. Similar textures but with $P \approx 15-16\text{ nm}$ occur when the sample is quenched from the $Col_t$ phase at 176°C, Fig. 12b. Another frequent texture is of wavy "ropes" that bend in space, revealing smooth terraces between them, Fig. 10c. The thickness of ropes varies from $3-4\text{ nm}$ to $5-6\text{ nm}$.

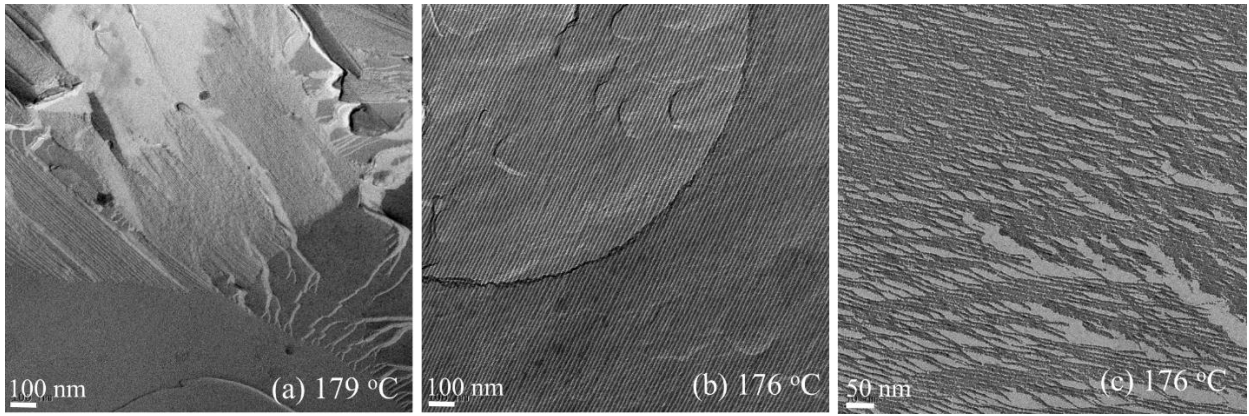

FIG. 12. Freeze-fracture TEM for the Col$_t$ phase of 1Cl-N(1,7)-O6 obtained by quenching the samples from (a) $T = 179°C$; (b,c) $T = 176°C$. The prevailing textures are linear periodic modulations (a,b) and wavy textures of "ropes" with bend deformations (c). The period is $7-8$ nm in (a) and $15-16$ nm in (b).

Unlike FFTEM, cryo-TEM produces direct images of ultrathin (up to 20 nm) films of the material on a holey carbon substrate. In the area of carbon support, the structure shows periodic ordering with the period that wary broadly from 3.8 nm to $7-8$ nm and even $12-15$ nm, Fig.13.

Periodic structures detected by TEM, combined with the optical observations of bend deformations, Fig. 6b, demonstrate that the positional ordering of Col$_t$ is of a columnar type. A layered smectic ordering is excluded since the optic axis in smectics experiences splay rather than bend. Broadly varying $P$ and formation of smooth terraces in FFTEM and cryo-TEM suggest that the interactions between columns are weak.

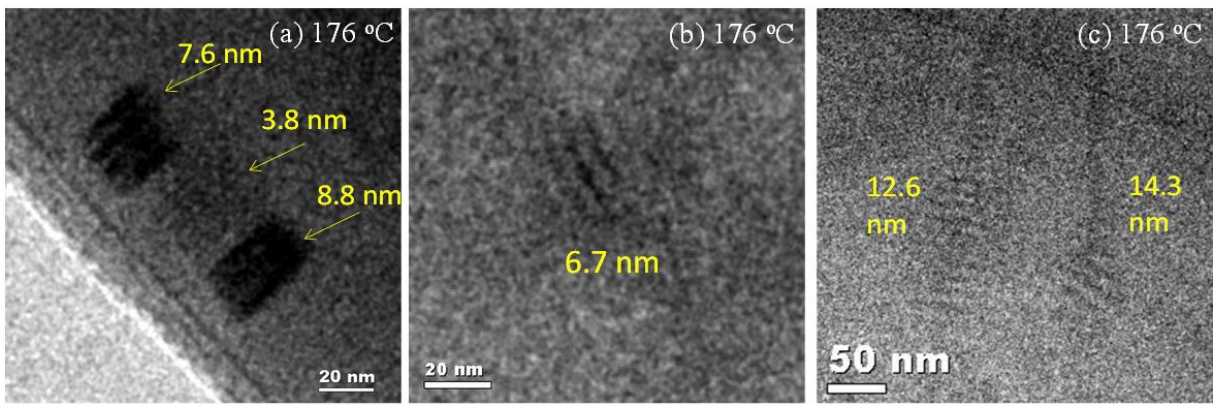

FIG. 13. Cryo-TEM textures of Col$_t$ phase of 1Cl-N(1,7)-O6 obtained by quenching from $T = 176°C$ with broadly varying period, from (a) 3.8 nm to (b) 6.7 nm and (c) $12-15$ nm.

## H. X-ray studies

Wide angle X-ray diffraction (WAXD) and Small Angle X-ray Scattering (SAXS) characterization of 1Cl-N(1,7)-O6 was performed for two types of samples: circular capillaries and flat cells aligned homeotropically by an AC electric field.

**Capillary sample**. The director is aligned perpendicularly to the incident X-ray beam by a magnetic field $\mathbf{B}$, Fig. 14. In both N and Col$_t$ phases, WAXS spectra show a broad peak in the equatorial plane (along the direction perpendicular to $\mathbf{B}$) at the scattering vector $q = M_{WAXS} = 14.2 \text{ nm}^{-1}$ which corresponds to a spacing $d_{WAXS} = 2\pi / M_{WAXS} = 0.45 \text{ nm}$. The correlation length associated with this spacing is very small, roughly 0.3 nm, as can be estimated from the inverse of the half-width-at-half-maximum, which implies that there is no translational periodicity. Because of this, the lateral spacing of 0.45 nm should be associated with the typical separations of molecular legs and their thicknesses. These legs are seen by the probing beam at different orientations; note that the basic element of the structure, a benzene ring, has a diameter of 0.45 nm. SAXS spectra depend strongly on the temperature. In the N phase, $T = 204°C$, there are broad meridional reflexes located along the line parallel to $\mathbf{B}$, at $M_1 = 1.73 \text{ nm}^{-1}$ ($d_1 = 3.63 \text{ nm}$). The distance $d_1 = 3.63 \text{ nm}$ is somewhat longer than the length $l_b \approx 3.0 - 3.2$ nm of the molecule measured along its bisector; it suggests that the molecules might associate into pairs with a slight shift along the director. As the temperature is lowered to $T = 180°C$, the peaks slightly shift to $d_1 = 3.51 \text{ nm}$.

The Col$_t$ phase is associated with new sharp peaks at $M_2 = 1.08 \text{ nm}^{-1}$ ($d_2 = 5.82 \text{ nm}$) and $M_3 = 1.52 \text{ nm}^{-1}$ ($d_3 = 4.1 \text{ nm}$). The ratio $M_3 / M_2 \approx 1.41 \approx \sqrt{2}$ signals tetragonal assembly, similarly to the results by Kang et al [18].

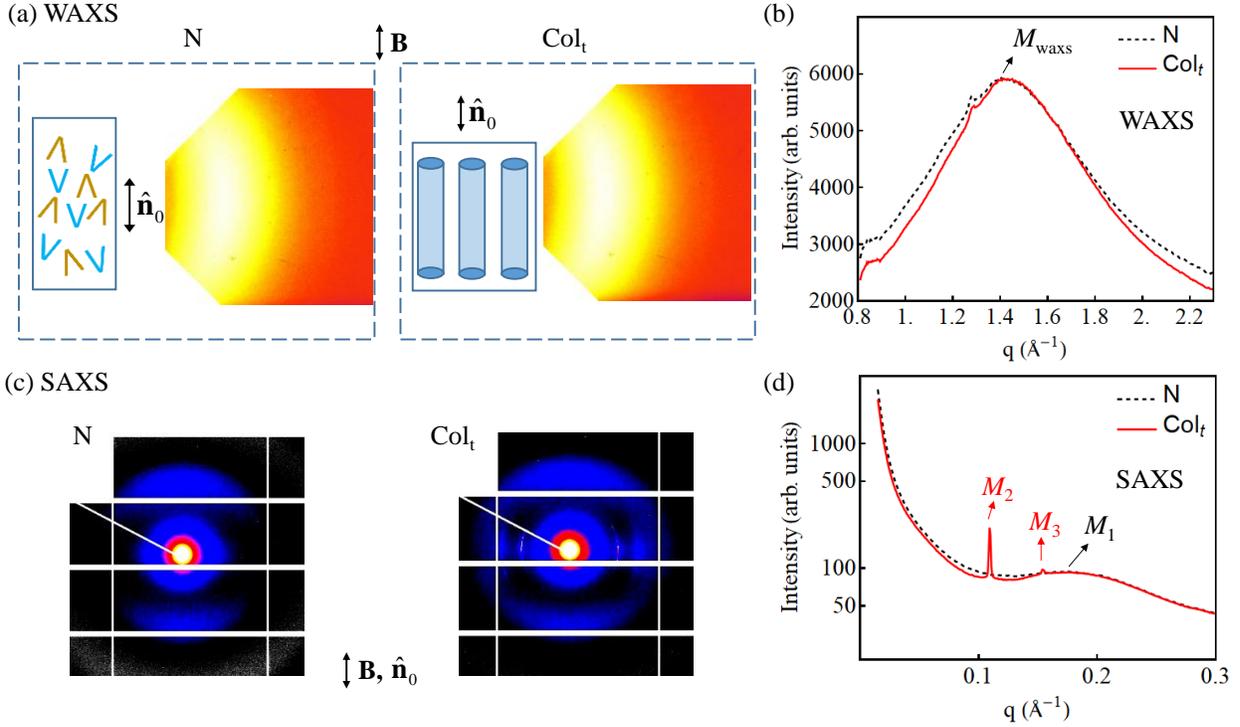

FIG. 14 X-ray diffraction data. (a) WAXS pattern of the N and Col$_t$ phases; (b) WAXS intensity vs. $q$ in the N and Col$_t$ phase; (c) SAXS pattern of the N and Col$_t$ phases; (d) SAXS intensity vs. $q$ in the N and Col$_t$ phases.

The detail analysis of SAXS pattern of the Col$_t$ phase is presented in Fig.15. The original SAXS pattern, Fig.14(c), contains the grid lines, which are the dead zones between the segments of the detector. To eliminate the effect of the grid, we assume that the pattern possesses the inversion symmetry with respect to the beam center and substitute the dead zones with the pixels from the opposite side, Fig.15(a). Both $M_2$ (5.8 nm) and $M_3$ (4.1 nm) reflexes exists at the direction perpendicular to the director, Fig. 15, which indicates an absence of positional order along the columns. However, the translational order of the columns themselves is strong, as indicated by the small widths of $M_2$, Fig. 15d, and $M_3$, Fig. 15f, reflexes in both radial and azimuthal directions. We estimate the correlations lengths to be 400 nm for $M_2$ and 300 nm for $M_3$ reflexes.

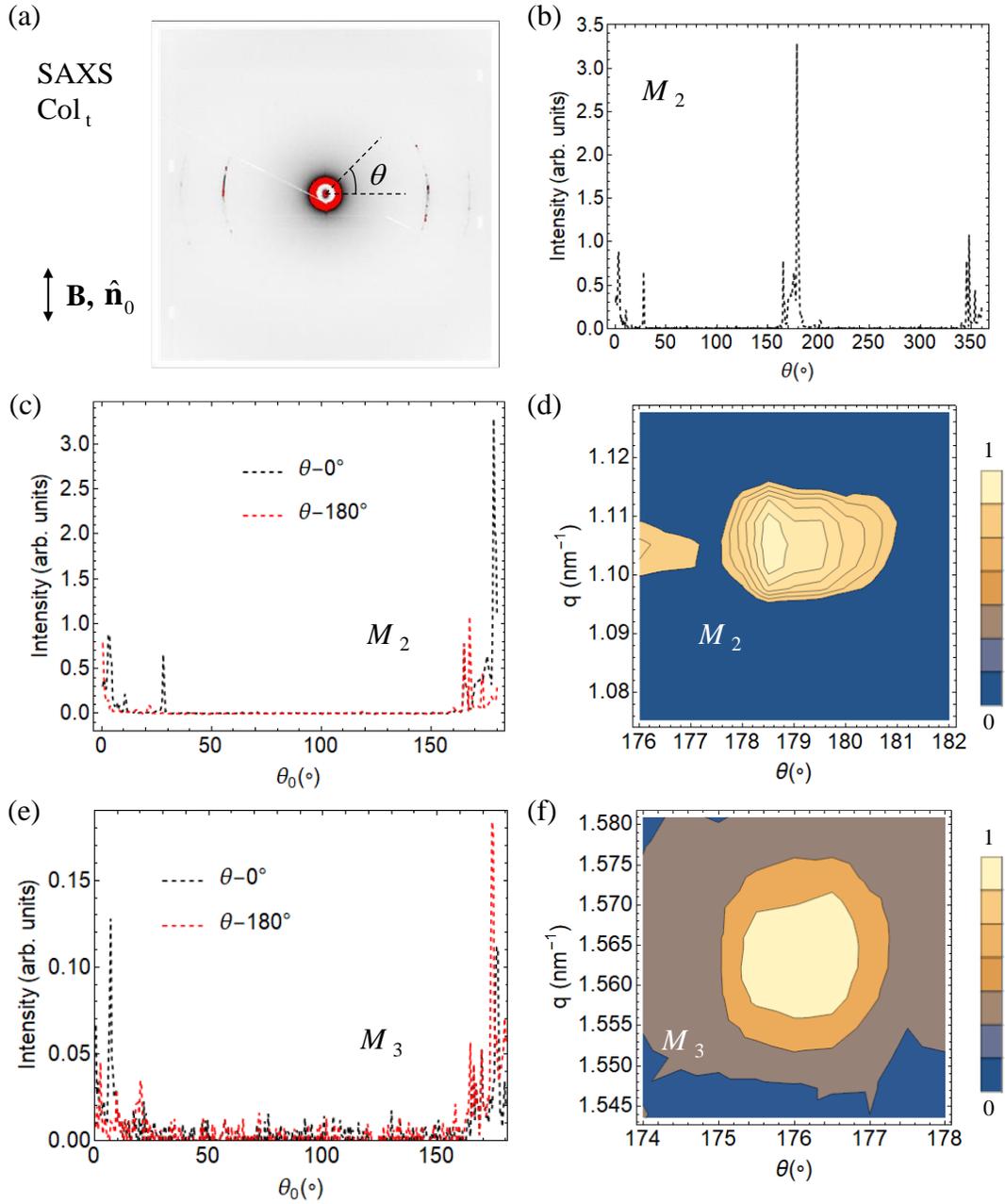

FIG. 15. Angular dependence of the SAXS signal in Capillary: (a) diffraction pattern; (b) signal intensity vs. the azimuthal angle $\theta$ for $M_2$ peak, where $2\pi/M_2 = 5.8\,\text{nm}$; (c) intensity vs. the angle $\theta_0$ for the $M_2$ peak, where $\theta_0 = \theta - 0°, 180°$, respectively; (d) intensity map for $M_2$ peak; (e) intensity vs. $\theta_0$ for $M_3$ peak, where $2\pi/M_3 = 4.1\,\text{nm}$; (f) intensity map for $M_3$ peak.

**Field aligned homeotropic Col$_t$ sample.** In this case, the director is oriented along the incident X-ray beam. The WAXS peak at $M'_{WAXS} = 14.2$ nm$^{-1}$ ($2\pi/M'_{WAXS} = 0.45$ nm), Fig. 16ab, and the two SAXS peaks, $M'_2 = 1.08$ nm$^{-1}$ ($2\pi/M'_2 = 5.82$ nm) and $M'_3 = 1.52$ nm$^{-1}$ ($2\pi/M'_3 = 4.1$ nm), Fig. 16cd, are the same as the peaks observed in the capillary.

The azimuthal analysis of the SAXS reflexes from the homeotropic sample confirms the tetragonal packing of columns, Fig. 17. Both $M'_2$ (5.8 nm) and $M'_3$ (4.1 nm) reflexes appear as sets of four peaks separated by 90 degrees, Fig.17b,c,e; different sets correspond to the different domains, developed because of the in-plane tetragonal orientation degeneracy. The strongest intensity peaks appear at 19° for $M'_2$ and 64° for $M'_3$, Fig. 17c,d,e,f, which means that there is an expected 45° difference between the two periodic arrangements. Note that $M'_2$ reflex, Fig. 17d, and $M'_3$ reflex, Fig. 17f, are more narrow than $M_2$, Fig. 17d, and $M_3$, Fig. 17f, revealing that under a homeotropic alignment, the correlation length of column packing is about 1 μm.

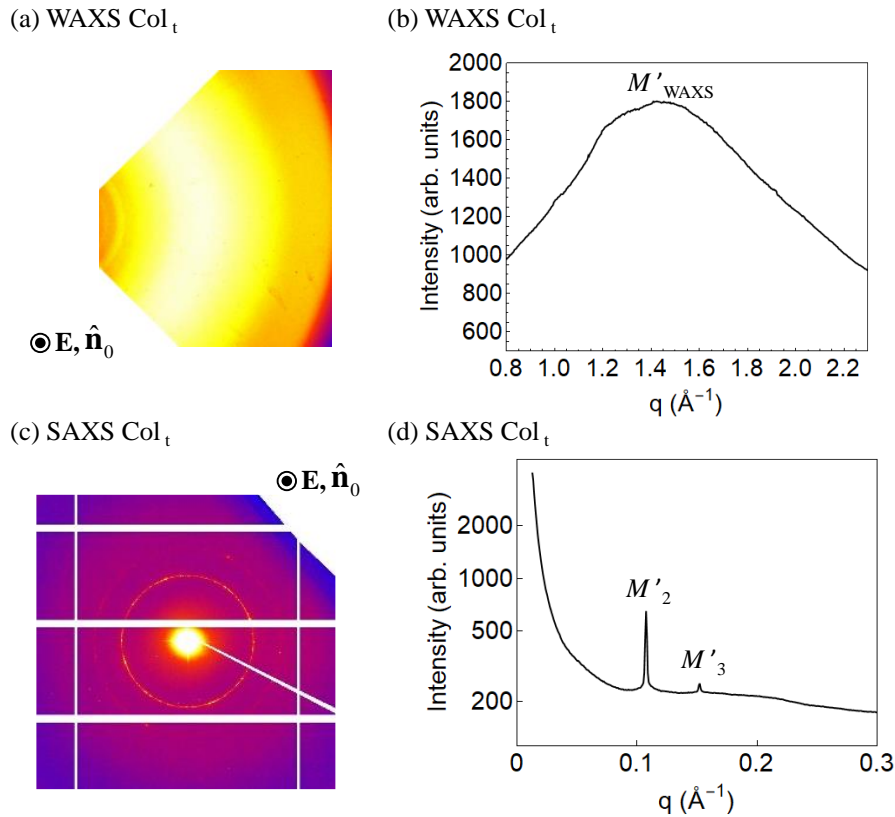

FIG. 16 X-ray diffraction at the sample aligned by the electric field at $T = 176$°C in the Col$_t$ phase. (a) WAXS pattern; (b) WAXS intensity vs. $q$; (c) SAXS pattern; (d) SAXS intensity vs. $q$.

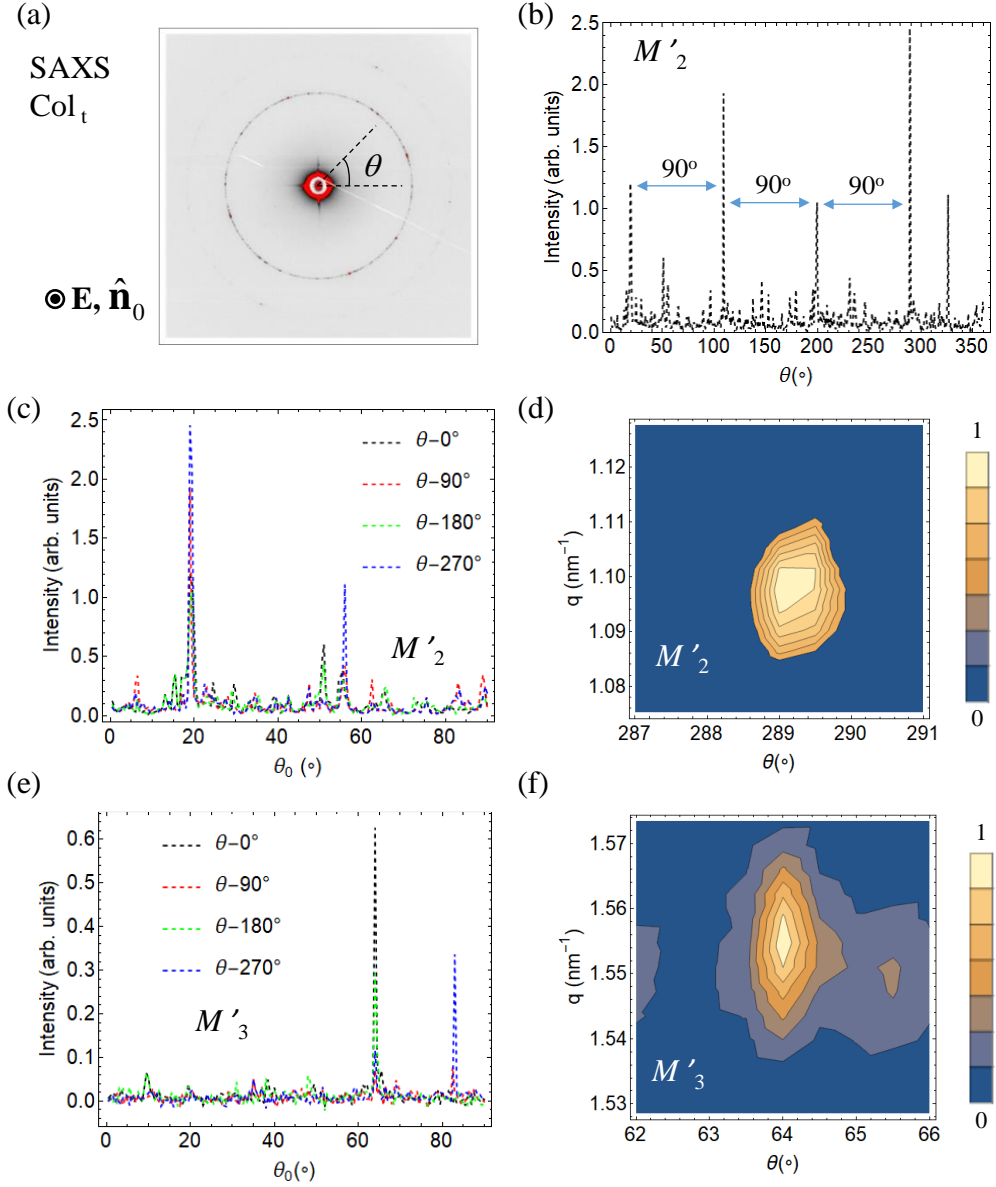

FIG. 17 Angular dependence of the SAXS signal: (a) diffraction pattern; (b) signal intensity vs. the azimuthal angle $\theta$ for $M'_2$ peak, where $2\pi/M'_2 = 5.8\,\text{nm}$; (c) intensity vs. the angle $\theta_0$ for the $M'_2$ peak, where $\theta_0 = \theta - 0°, 90°, 180°, 270°$, respectively; (d) intensity map for $M'_2$ peak; (e) intensity vs. $\theta_0$ for $M'_3$ peak, where $2\pi/M'_3 = 4.1\,\text{nm}$; (f) intensity map for $M'_3$ peak.

Based on these results, we propose that the Col$_t$ phase is a tetragonal packing of columns of two types, one polar and one apolar, Fig.18. In the cross-section of each column, one finds two ABC molecules, the $\lambda$-planes of which are orthogonal to each other. These two molecules can point in the same direction, Fig.18 a,b, or in the opposite directions, Fig.18c. A column formed by the pairs of the same orientation is polar, with polarization vector either "up" (yellow columns in Fig.18d) or down (blue columns in Fig.18d). Columns formed by stacks of the apolar pairs are apolar themselves, and marked as green in Fig. 18d. The polar columns form an antiferroelectric lattice, in which any "up" column is neighbored by four "down" columns and vice versa. The polar character of the columns is essential as it allows the system to avoid hexagonal packing and to form a square lattice instead. The cross-section of each column is of a four-fold symmetry. The four arms of all polar columns are parallel to each other, while the four arms of the apolar columns are oriented by $45°$ with respect to the polar ones. The X-ray measured periodicity of 5.8 nm is associated with the distances between columns of the same polarity along the sides of a square lattice, while the 4.1 nm period with the distances along the diagonal direction. The correlation lengths corresponding to $M_2$ and $M_3$ reflexes are both about 600 nm. Note that the columnar phase formed by ABC molecules is very different from the columnar phase observed in disk like molecules because in the Col$_t$ phase molecular arms are almost parallel rather than perpendicular to the axes of columns.

The chemical composition of the studied material is $C_{64}H_{56}Cl_2N_2O_{10}$, and the molecular weight is $m = 1084$ amu $= 1.800 \times 10^{-24}$ kg. Assuming the mass density in the Col$_t$ phase $\rho_a \sim 10^3$ kg/m$^3$ which is typical for liquid crystals [44], one can estimate that a volume $V = 5.8$ nm $\times 5.8$ nm $\times 1$ nm $= 3.364 \times 10^{-26}$ m$^3$ should contain $n = 18.69$ ABC molecules. In the proposed packing, the number of molecules in the elementary 5.8 nm $\times$ 5.8 nm square is 8, thus the molecular separation along $z$-axis is $d_z = 0.43$ nm, which is a very reasonable result. The number of molecules in an elementary square cannot be reduced below 8 because for density $\rho_a \sim 10^3$ kg/m$^3$, $d_z$ would be prohibitively small.

The model in Fig.18 places all the molecular pairs in the same plane of view. This is done for clarity only, as in the real structure this restriction is absent: the molecules that align in the opposite direction have the centers of mass shifted along the columnar axis. The polar and apolar columns can be arranged in a variety of ways with each other. In Fig.18, we show four possible local structures that show some of the possible geometries. A unifying theme of these arrangements is an antiferroelectric order of the polar columns with the gaps filled by apolar columns. The distinction comes from how the molecules in apolar columns contact the molecules in the polar ones. These contacts can be locally "parallel" (molecules from the neighbouring columns are of the same color) or "anti-parallel" (the neighbouring molecules are of the different colors). Note that the transition from one type of arrangement to another in Fig.18d,e,f,g, requires only a simple 90-degree rotation of apolar columns. One might expect that all variations of polar-apolar columns arrangements are possible when the system is cooled down from the nematic phase into the $Col_t$ phase, since it would increase the entropy of packing. The possibility of different geometrical arrangements between the polar and apolar columns is probably the for multiple periodicities revealed by TEM observations in Figs. 12,13.

An important feature of the proposed model of the columnar phase in Fig.18 is that the plane of the ABC molecules is parallel to the columnar axes. This arrangement, supported by the X-ray, optical, dielectric and electro-optical data, is very different from the conventional columnar phases formed by disk like molecules that stack on top of each other, with their planes being perpendicular to the columns.

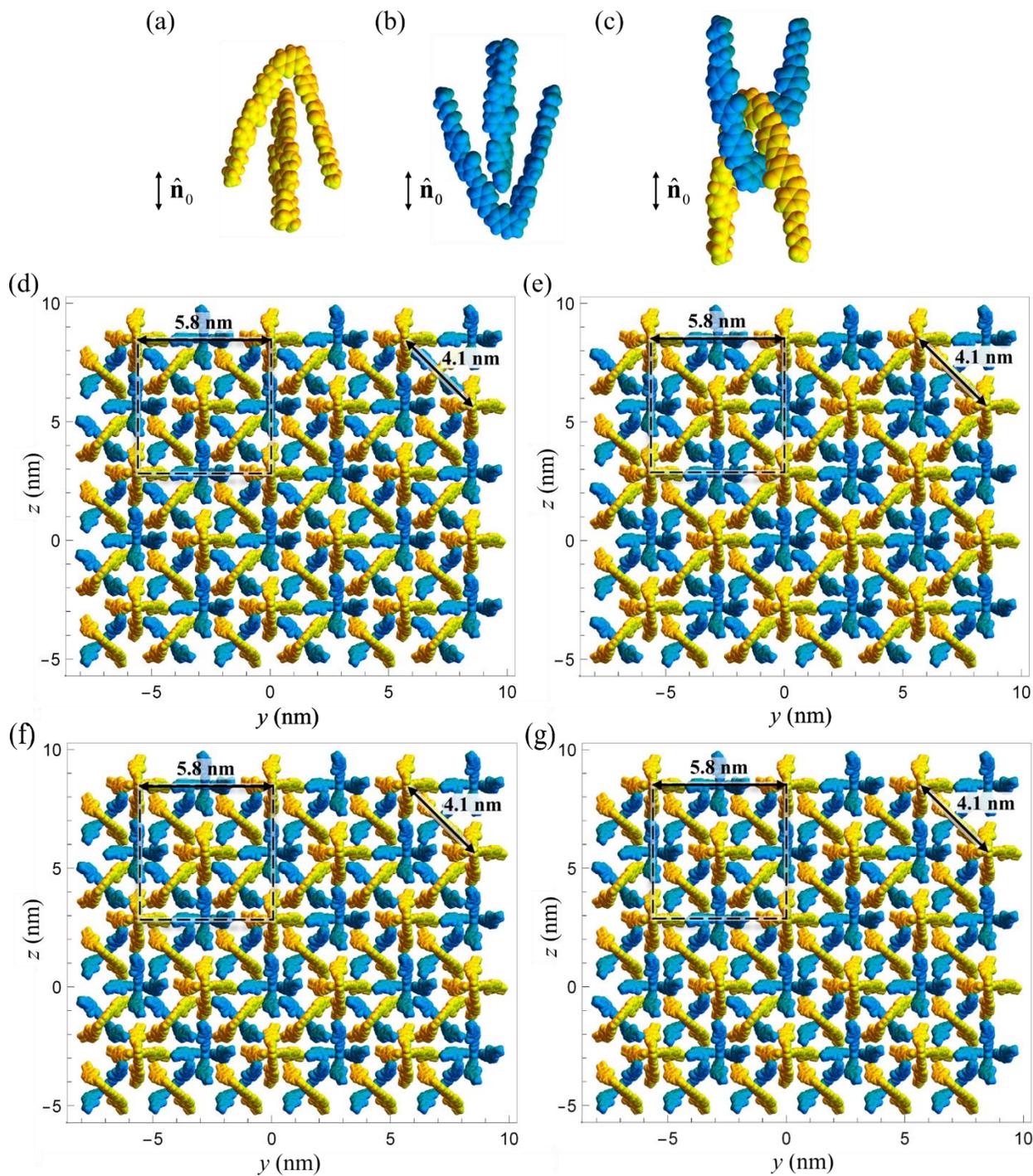

FIG. 18. Columnar packings in Col$_t$ phase: (a) two ABC molecules point up (b) two ABC molecules point down (c) two molecules point in opposite directions; (d, e, f, g) Intercalated assembly of polar (a,b) and apolar (c) columns into tetragonal structures of Col$_t$.

## IV. DISCUSSION and CONCLUSION

The mesomorphism of ABC-compounds with 1,7-naphthalene central core was reported by Lee et al [45]. This family of materials exhibits a broad variety of phases [17, 45]. Especially interesting phase sequence was suggested for the studied 1Cl-N(1,7)-O6 [18], with the N phase being accompanied at lower temperatures by a tetragonal packing of cylinders with double-twisted director. In the discussion below, we first address the properties of the N phase and then the properties of the $Col_t$ phase.

Homeotropic textures of the N phase and the fact that its birefringence is positive suggest that the director is defined by the bisectors of each $\lambda$-shaped ABC molecule. The symmetry of a single molecule is $C_{2V}$, which is lower than the macroscopic nematic symmetry $D_{\infty h}$. The molecules are thus expected to rotate around the direction close to the bisector and flip-flop between "up" and "down" orientations. The measured values for twist $K_{22} \sim 5 \text{ pN}$ and bend $K_{33} \sim 15 \text{ pN}$ moduli are close to their counterparts in the rod-like nematics. The splay constant is anomalously low $K_{11} \sim 2 \text{ pN}$, which we associate with the flip-flops of the $\lambda$-shaped ABC molecules that accommodate the splay easily. The inequalities $K_{11} < K_{22} < K_{33}$ in the ABC nematic phase are very different from the trends $K_{22} < K_{11} < K_{33}$ and $K_{33} < K_{22} < K_{11}$ in rod-like and OBC nematics, respectively. It would be of interest to explore whether the anomalous elastic properties of ABC materials extend also to the saddle-splay deformations and to the "biaxial splay" recently discussed by J. Selinger [46].

The appearance of a low-temperature $Col_t$ phase is associated with the gradual freezing of the flip-flops and appearance of the positional order of tetragonal symmetry. The trend is evidenced already in the N phase, as $K_{11}$ increases as the temperature decreases, which can be associated with the formation of molecular dimers such as shown in Fig. 18c. This temperature behavior differs from that of the bend modulus $K_{33}$ upon approaching the nematic to twist-bend nematic transition [7, 9, 14, 25,26]. In the latter case, the molecules acquire a more bent shape at lower temperatures and $K_{33}$ decreases, except in the very vicinity of the transition.

The $Col_t$ phase is of a tetragonal symmetry. The X-ray diffraction data suggests that this phase is formed by columnar aggregates of diameter roughly equal the distance between the legs of a single $\lambda$-shaped molecule. The polar columns are of alternating up and down polarities. The director, being parallel to the columns, is unidirectional in the ground state, but can be bent easily. Equidistance of columns makes splay and twist deformations difficult. The typical distances between columns of the same polarity are 5.8 nm and 4.1 nm. Apparently, the cohesive forces between the columns are weak, as the TEM study show a broad spectrum of periods, including the values close to distances 5.8 nm and 4.1 nm measured by X-ray scattering. Apolar (on average) antiferroelectric character of packing explains relatively weak dielectric permittivity of the material.

## ACKNOWLEDGMENTS

We thank S. Kang and J. Watanabe for the supply of material for the initial stage of this study. This research used the 11-BM CMS beamline of National Synchrotron Light Source-II (NSLS-II), Brookhaven National Laboratory (BNL), a U.S. Department of Energy User Facility operated for the Office of Science by BNL under Contract DE-SC0012704. FFTEM and cryo-TEM studies were performed at the (cryo) TEM facility at the Advanced Materials and Liquid Crystal Institute, Kent State University, supported by the Ohio Research Scholars Program Research Cluster on Surfaces in Advanced Materials. The work was supported by NSF grant DMR-1905053.

**Supplementary materials for**

**"Liquid crystal phases with unusual structures and physical properties formed by acute-angle bent-core molecules"**

**Supplementary Figure**

**Synthesis of Highly Bent Liquid Crystal**

**Supplementary Figure**

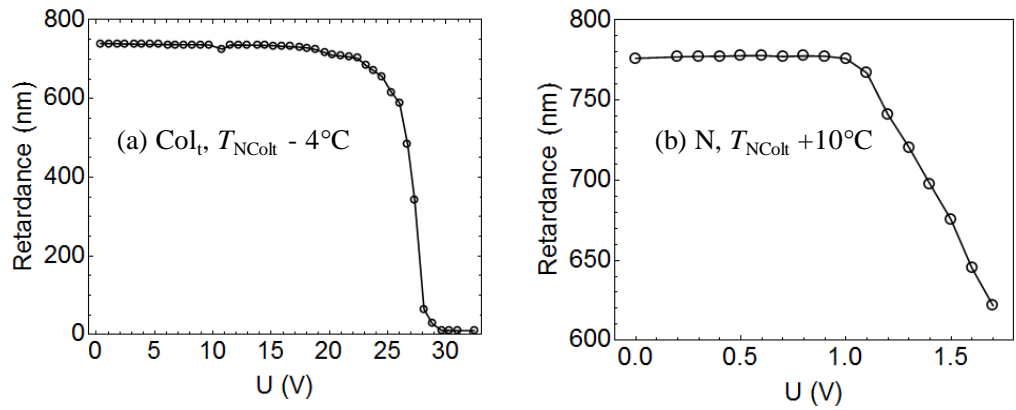

Fig. S1. Voltage dependence of optical retardance in the $Col_t$ (a) and N (b) phases at 1 kHz.

# Synthesis of Highly Bent Liquid Crystal

## 1. Materials and methods

All reagents and solvents were available commercially and used as received unless otherwise stated. $^1$H and $^{13}$C NMR spectra were recorded on a Bruker 400 MHz NMR spectrometer using CDCl$_3$ as solvent. Chemical shifts are in δ unit (ppm) with the residual solvent peak as the internal standard. The coupling constant (*J*) is reported in hertz (Hz). NMR splitting patterns are designed as follows: s, singlet; d, doublet; t, triplet; and m, multiplet. Column chromatography was carried out on silica gel (230-400 mesh). Analytical thin layer chromatography (TLC) was performed on commercially coated 60 mesh F$_{254}$ glass plates. Spots on the TLC plates were rendered visible by exposing to UV light.

## 2. Synthesis of the intermediates and target

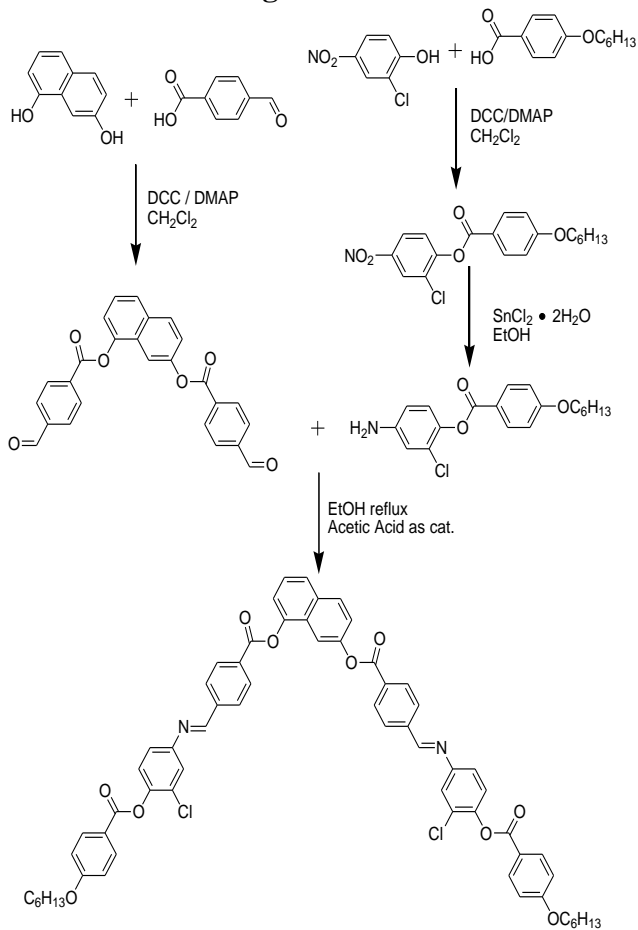

Fig. S4.  Synthetic route of the highly bent liquid crystal

## 3. Synthesis details and NMR data

### 3-choloro-4-(4-hexyloxybenzoyl)-nitrobenzene

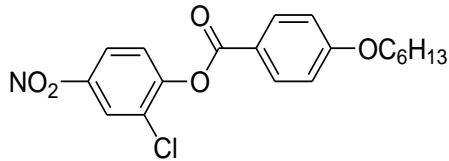

To a solution of 2-chloro-4-nitrophenol (1.73 g, 10 mmol) and 4-hexoxybenzoic acid (2.23 g, 10 mmol) in dichloromethane (DCM, 50 mL) was added dicyclohexylcarbodiimide (DCC, 2.07 g, 10 mmol) and 4-dimethylaminopyridine (DMAP as catalyst, 10 mg). The solution was stirred 24 hours, filtered, and purified by silica gel column chromatography to give the product as white solid (2.8 g, 74%). $^1$H NMR (400 MHz, CDCl$_3$): δ= 8.48 (d, $J$ = 2.63 Hz, 1H), 8.22 (m, 1H), 8.15 (d, $J$ = 8.10 Hz, 2H), 7.51 (d, $J$ = 8.92 Hz, 1H), 7.00 (d, $J$ = 8.92 Hz, 2H), 4.06 (t, $J$ = 6.52 Hz, 2H), 1.83 (m, 2H), 1.48 (m, 2H), 1.37 (m, 4H), 0.92 (t, $J$ = 6.89 Hz, 3H).

### 3-choloro-4-(4-hexyloxybenzoyl)oxyaniline

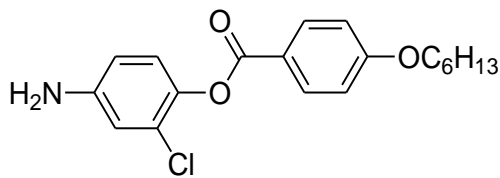

The nitro compound (3.23 g, 8.54 mmol) and Tin (II) chloride dihydrate (9.66 g, 42.7 mmol) were added to ethanol and reflux for 24 hours, after workup, the product was purified by silica gel column chromatography to give a white solid (2.38 g, 80 %). $^1$H NMR (400 MHz, CDCl$_3$): δ= 8.16 (d, $J$ = 8.92 Hz, 2H), 7.02 (d, $J$ = 8.65 Hz, 1H), 6.95 (d, $J$ = 8.91 Hz, 2H), 6.77 (d, $J$ = 2.61 Hz, 1H), 6.60 (d, $J$ = 8.61 Hz, 1H), 4.06 (t, $J$ = 6.55 Hz, 2H), 3.70 (s, 2H), 1.83 (m, 2H), 1.46 (m, 2H), 1.38-1.34 (m, 4H), 0.92 (t, $J$ = 6. 97 Hz, 3H).

### 1,7-naphthylene bis(4-formylbenzoate)

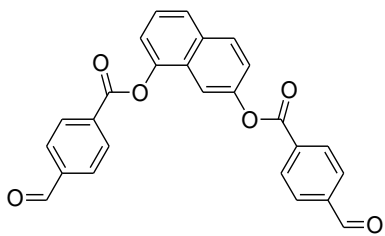

1,7-dihydroxy naphthalene (0.57 g, 3.55 mmol) and 4-formyl benzoic acid (1.12 g, 7.46 mmol) was dissolved in DCM, then dicyclohexylcarbodiimide (DCC, 1.54 g, 7.46 mmol) and 4-dimethylaminopyridine (DMAP, 10 mg) were added. After 24 hours stirring, the product was filtered, and purified by flash column chromatography to give a white solid (0.98 g, 65%). $^1$H NMR (400 MHz, CDCl$_3$): δ = 10.16 (s, 1H), 10.14 (s, 1H), 8.48 (d, $J$ = 8.23 Hz, 2H), 8.37 (d, $J$ = 8.24 Hz, 2H), 8.07 (d, $J$ = 8.36 Hz, 1H), 8.01 (m, 3H), 7.87 (d, $J$ = 8.26 Hz, 1H), 7.74 (d, $J$ = 2.26 Hz, 1H), 7.57 (t, $J$ = 7.93 Hz, 1H), 7.44 (m, 2H). $^{13}$C NMR (100 MHz, CDCl$_3$): δ = 191.99, 164.79, 164.64, 149.73, 146.95, 140.39, 140.27, 133.53, 131.42, 130. 38, 130.22, 126.87, 126.22, 122.54, 119.71, 113.00.

**1,7-naphthylene bis(4-(3-chloro-4-(4-(hexyloxy)benzoyloxy) phenyliminomethyl)benzoate)**

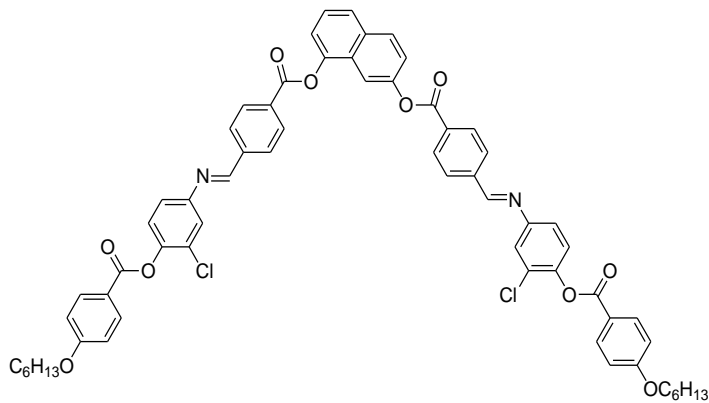

A mixture of 1,7-naphthylene bis(4-formylbenzoate) (0.30 g, 0.7 mmol), 3-choloro-4-(4-hexyloxybenzoyl)oxyaniline (0.485 g, 1.43 mmol) were dissolved in ethanol and then a few acetic acid was added. The solution was refluxed under N$_2$ atmosphere for overnight. The mixture was recrystallized from chloroform/ethanol twice to give a yellow solid (0.55 g, 72 %). $^1$H NMR (400 MHz, CDCl$_3$): δ = 8.57 (s, 2H), 8.55 (s, 2H), 8.44 (d, $J$ = 8.29 Hz, 2H), 8.33 (d, $J$ = 8.29 Hz, 2H), 8.18 (d, $J$ = 8.90 Hz, 4H), 8.10 (d, $J$ = 8.29 H, 2Hz), 8.05 (d, $J$ = 8.35 Hz, 2H), 8.03 (d, $J$ = 9.14 Hz, 1H), 7.86 (d, $J$ = 8.35 Hz, 1H), 7.78 (d, $J$ = 2.26 Hz, 1H), 7.57 (t, $J$ = 7.80 Hz, 1H), 7.48-7.44

(m, 2H), 7.40 (t, *J* = 2.80 Hz, 2H), 7.33-7.31 (m, 2H), 7.25-7.21 (m, 2H), 7.00 (d, *J* = 8.84 Hz, 4H), 4.06 (t, *J* = 6.52 Hz, 4H), 1.83 (m, 4H), 1.46 (m, 4H), 1.38-1.34 (m, 8H), 0.92 (t, *J* = 6.89 Hz, 6H). $^{13}$C NMR (100 MHz, CDCl$_3$): δ = 165.27, 165.09, 164.71, 164.40, 160.35, 150.34, 150.07, 149.83, 147.08, 146.37, 140.97, 140.81, 132.46, 133.14, 132.38, 132.21, 131.41, 131.28, 130.52, 129.72, 129.56, 128.09, 126.13, 123.07, 122.70, 119.71, 114.99, 113.18, 68.95, 32.13, 29.63, 26.25, 23.19, 14.63.